\RequirePackage{fix-cm}
\documentclass[twocolumn]{svjour3}          % twocolumn
\smartqed  % flush right qed marks, e.g. at end of proof

\usepackage{times}
\usepackage{url}
\usepackage{amsmath}
\usepackage{natbib}
\usepackage{threeparttable}
\usepackage[dvipdfmx]{graphicx}
\usepackage[misc,geometry]{ifsym}
\usepackage{booktabs}

\makeatletter
\let\MYcaption\@makecaption
\makeatother

\usepackage{subcaption}
\makeatletter
\let\@makecaption\MYcaption
\makeatother

\journalname{Social Network Analysis and Mining}
\begin{document}

\title{Empirical Evaluation of Link Deletion Methods\\for Limiting Information Diffusion on Social Media%\thanks{Grants or other notes
}


\author{Shiori Furukawa     \and
        Sho Tsugawa %etc.
}


\institute{S.  Furukawa \Letter \at
Graduate School of Engineering,
Information and Systems, University of Tsukuba\\
 1--1--1 Tennodai, Tsukuba, Ibaraki 305--8573, Japan \\
              Tel./Fax: +81-29-853-5597\\
              \email{s.furu@mibel.cs.tsukuba.ac.jp}           %  \\
           \and
           S. Tsugawa\at
Faculty of Engineering, Information and Systems, University of Tsukuba\\
              \email{s-tugawa@mibel.cs.tsukuba.ac.jp}
}

\date{Received: date / Accepted: date}

\maketitle

\begin{abstract} % abstract
%\parttitle{First part title} %if any
Although beneficial information abounds on social media, the
 dissemination of harmful information such as so-called ``fake news''
 has become a serious issue. Therefore, many researchers have devoted
 considerable effort to limiting the diffusion of harmful information. A
 promising approach to limiting diffusion of such information is link
 deletion methods in social networks. Link deletion methods have been
 shown to be effective in reducing the size of information diffusion
 cascades generated by synthetic models on a given social network. In
 this study, we evaluate the effectiveness of link deletion methods by
 using actual logs of retweet cascades, rather than by using synthetic
 diffusion models. Our results show that even after deleting 10\%--50\% of
 links from a social network, the size of cascades after link deletion
 is estimated to be only 50\% the original size under the optimistic estimation, which suggests that the effectiveness of the link deletion strategy for suppressing information diffusion is limited. Moreover, our results also show that there is a considerable number of cascades with many seed users, which renders link deletion methods inefficient.
\end{abstract}
\maketitle
\section{Introduction}
\label{sec:introduction}

An abundance of beneficial information is available on social media~\citep{fan2013least}.
Social media users can disseminate information posted by other users through functionalities such as retweeting on Twitter and sharing on Facebook~\citep{doerr2012rumors}.
Through such word-of-mouth information diffusion, information about important breaking news and newly created useful content is disseminated quickly and widely on social media.
Because social media users can access such useful information, social media is considered to be an important platform for both those who produce information and those who consume it.

In contrast, the dissemination of harmful information, including so-called ``fake news'', misinformation, and hate speech, has become a serious issue~\citep{kimura2008minimizing,Sharma19:Combating,Mathew19:Spread}.
For example, during the coronavirus disease 2019 (COVID-19) outbreak, there was widespread dissemination of misleading and false information on social media, resulting in people believing in false theories about the disease and the virus that causes it~\citep{jolley2020pylons}.
In addition, it has been reported that an increasing number of people are relying on misleading information on social media and as a result are hesitant to get vaccinated without first seeking professional advice~\citep{van2020inoculating}.

Many researchers have devoted substantial effort to limiting the
diffusion of harmful information on social
media~\citep{perez-rosas-etal-2018-automatic,shah2011rumors,leskovec2009meme,Sharma19:Combating}.
\cite{perez-rosas-etal-2018-automatic}
proposed a method to automatically detect fake news on social media.
\cite{shah2011rumors} proposed a method to estimate the users who are
the source of rumors on social media.  \cite{leskovec2009meme} proposed a framework for tracking the spread of fake news.

\begin{figure}[t]
\begin{minipage}{.99\columnwidth}
\begin{center}
\includegraphics[bb=0 0 1550 1150,clip,scale=0.1]{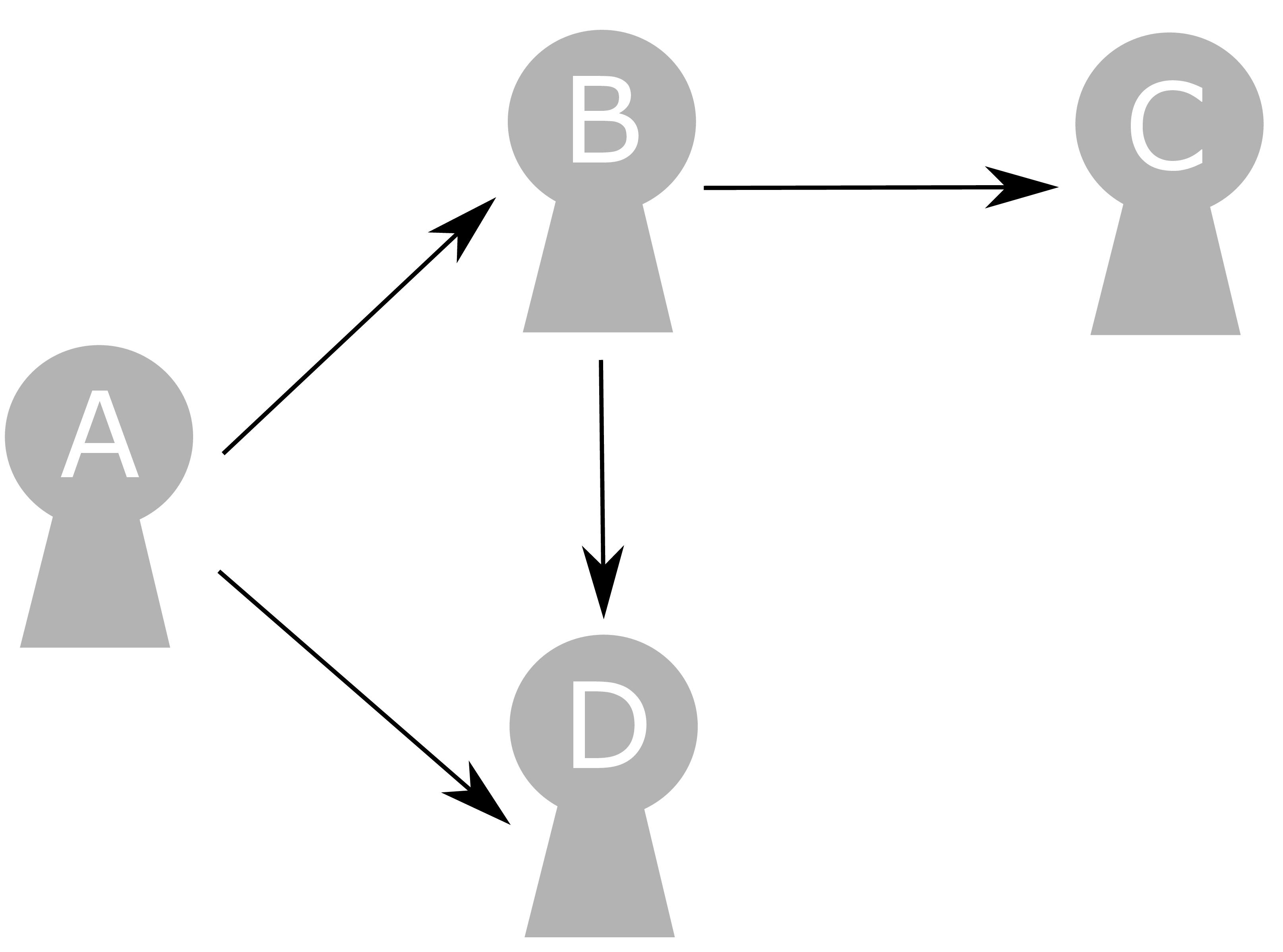}
\subcaption{Information flows before blocking information diffusion. Information posted by user A spreads to user A's followers, namely, users B and D. The information is reposted by user B, then the information spreads to users C and D.}
\label{fig:human_diffusion_before}
\end{center}
\end{minipage} \\
\begin{minipage}{.99\columnwidth}
\begin{center}
\includegraphics[bb=0 0 1550 1150,clip,scale=0.1]{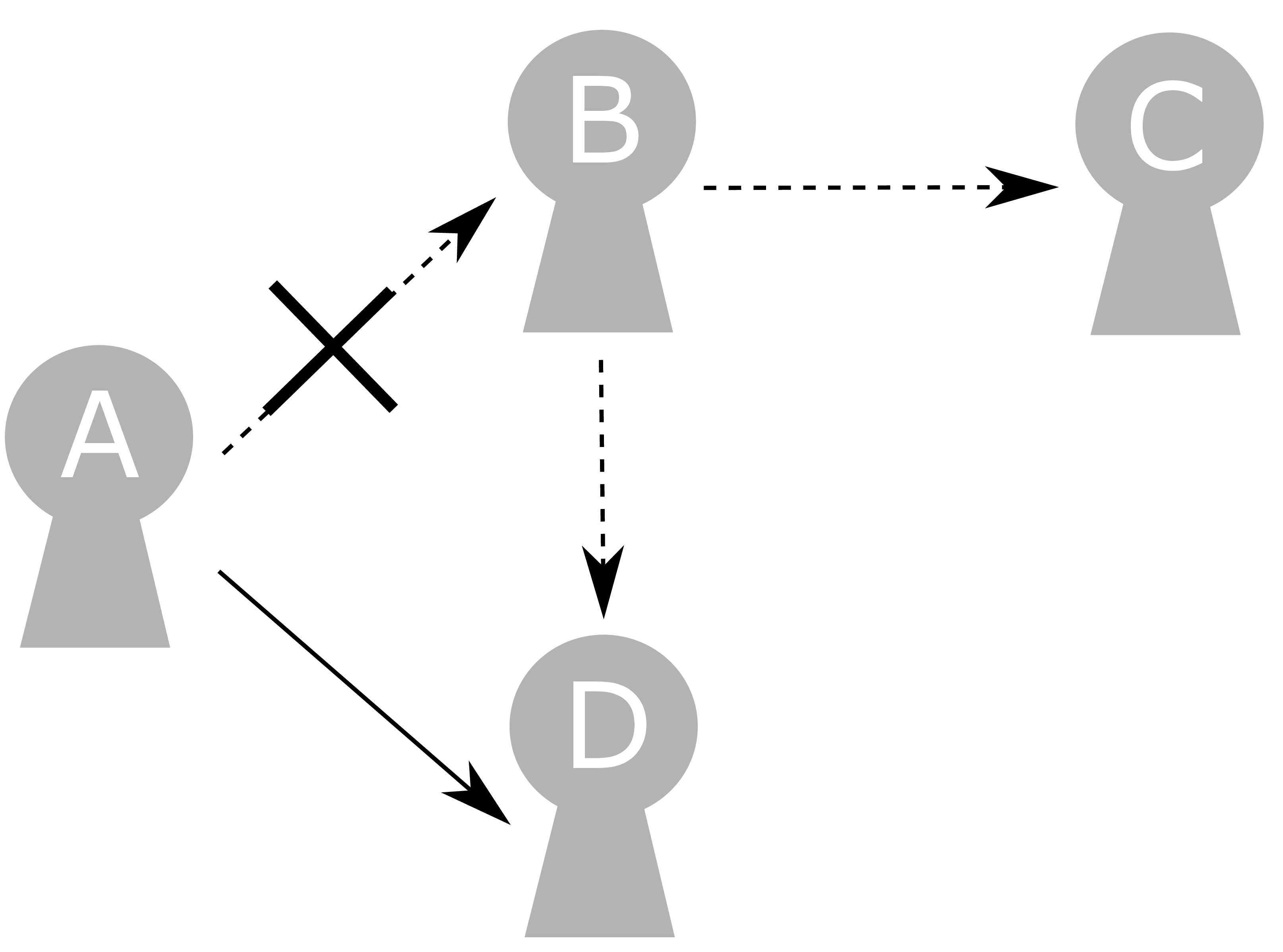}
\subcaption{Information flows after blocking information diffusion. When information diffusion from user A to user B is blocked, user B cannot repost the information. As a result, user C cannot receive information from user B.}
\label{fig:human_diffusion_after}
\end{center}
\end{minipage}
\caption{Example of blocking information diffusion between users}
\label{fig:diffusion}
\end{figure}

Methods to block information diffusion between (a small number of) specific users have been actively studied as promising approaches to limiting harmful information diffusion~\citep{yan2019rumor,kimura2008minimizing,tong2012gelling,10.1007/978-3-662-47401-3_9}.
Blocking one user's information from spreading to other users can reduce the final size of information diffusion cascades.
Figure~\ref{fig:diffusion} shows an example of blocking information diffusion between users.
As shown in Fig.~\ref{fig:diffusion}, by blocking information diffusion between only users A and B, the number of users who finally receive the information will decrease from 4 to 2.
However, if information diffusion is blocked too heavily, there is a risk of blocking diffusion of useful information.
Therefore, it is desirable to control the diffusion of harmful information by blocking the diffusion of information between as few users as possible.

The problem with limiting the size of information diffusion cascades by blocking information diffusion between a limited number of users is formulated as a link deletion problem in social networks~\citep{kimura2008minimizing}.
A social network is defined as a graph $G=(V,E)$, in which social media users are represented by nodes and relationships among them are represented by links~\citep{10.1145/502512.502525}.
Information posted by nodes in a social network spreads through links between nodes.
Given a social network $G$ and an integer $k$, the link deletion problem aims to determine the combination of $k$ links that should be removed in order to minimize the size of information diffusion cascades on social network $G$.
Many studies have proposed various link deletion methods to control the size of information diffusion cascades~\citep{kimura2008minimizing,tong2012gelling,khalil2014scalable,yan2019rumor}.
Previous studies have used synthetic information diffusion models to evaluate the effectiveness of the link deletion methods~\citep{tong2012gelling,yan2019rumor,ZAREIE2022100206}.
For example, \cite{tong2012gelling} evaluated the effectiveness of link deletion methods in reducing the size of information diffusion using a popular synthetic diffusion model called the susceptible-infectious-susceptible (SIS) model.

In this study, we aim to reveal the benefits and limitations of link deletion in limiting the spread of information diffusion on {\em real} social media.
Contrary to previous studies that used synthetic models, we utilize the actual log data to evaluate the effectiveness of link deletion methods.
Synthetic information diffusion models cannot fully reproduce the characteristics of information diffusion on actual social media~\citep{10.1145/1871437.1871737}.
Therefore, even if the link deletion method is effective in controlling the size of information diffusion generated by the synthetic information diffusion model, its effectiveness in controlling actual information diffusion on social media remains unclear.
We examine the size of information diffusion cascades when it is possible to block information diffusion between a certain number of user pairs identified using link deletion methods.
Because it is not easy to conduct an experiment that actually blocks information diffusion, we propose a method to estimate the size of retweet diffusion cascades after link deletion from a who-follows-whom network users and log data of information diffusion among them.
Based on the proposed method, we examine the effectiveness of link deletion methods including NetMelt~\citep{tong2012gelling}, a popular link deletion method, and other heuristics for limiting the size of information diffusion cascades on social media~\citep{KASHYAP2019631}.

This paper is an extended version of our previous conference short paper~\citep{10.1145/3487351.3488351}. We have conducted experiments on two additional datasets, and two additional link deletion methods for comprehensive examination of the effectiveness of link deletion methods. Moreover, we propose a new method to estimate the size of diffusion cascade after link deletion.  In our previous paper, we only evaluated the effectiveness of link deletion under the worst-case scenario.  In this paper, we evaluate  the effectiveness of link deletion both under the worst-case and more optimistic scenarios.

The contributions of the present study are summarized as follows.
\begin{itemize}
  \item We use four large-scale datasets of cascades rather than synthetic cascade models to evaluate the effectiveness of link deletion. We propose a method to estimate the size of a given cascade after link deletion. Using the proposed method, we examine the benefits and limitations of link deletion methods when applying them to actual social media.

  \item Our results suggest that there are limitations to link deletion
	methods in limiting the size of retweet cascades on
	social media. While existing
	studies~\citep{tong2012gelling,yan2019rumor} have demonstrated
	the effectiveness of link deletion methods in limiting the size
	of cascades generated by synthetic models, our results suggest
	that link deletion methods are not so effective in limiting the
	size of real diffusion cascades on social media both under the
	pessimistic and optimistic scenarios.

  \item We show that most of the cascades examined in this study have multiple sources, which makes limiting their sizes by link deletion difficult.
\end{itemize}

The remainder of this paper is organized as follows.
In Section~\ref{sec:related_work}, we introduce related work.
In Section~\ref{sec:diffusion}, we propose a method to estimate the size of cascades when link deletion methods are applied.
In Section~\ref{sec:method}, we explain the datasets and methodologies for the analyses.
In Section~\ref{sec:result}, we present the results and discuss their implications.
Finally, in Section~\ref{sec:conclusion}, we summarize this study and discuss future work.

\section{Related Work}
\label{sec:related_work}
Many researchers have conducted research to detect harmful information on social media~\citep{perez-rosas-etal-2018-automatic,shah2011rumors,10.1145/3386253,Helmstetter18:Weakly}.
\cite{perez-rosas-etal-2018-automatic} proposed a method to automatically detect fake news on social media.
They analyzed actual fake news texts and developed a model to predict whether a given news text is fake news by combining information on vocabulary, syntax, and semantics that are frequently used in actual fake news.
\cite{10.1007/978-3-030-59430-5_3} proposed an approach to detect non-factual twitter accounts.
\cite{shah2011rumors} proposed a method to estimate the source of rumors using maximum likelihood estimation.
To control the spread of harmful information, both detecting harmful information and limiting its spread are required~\citep{Sharma19:Combating}.
While the above mentioned studies focused on the former, the present study focuses on the latter.

To limit the spread of information diffusion on a social network, several studies have proposed methods to change the network topology of the given network so that the size of the spread of information diffusion cascades on the network are minimized~\citep{khalil2014scalable,kimura2008minimizing}.
Some of these studies have proposed removing several nodes from the network~\citep{wang2013negative,Alorainy2022}.
%In particular, previous studies have shown that removing high-out-degree nodes is an effective strategy~\citep{CRUCITTI2004388,PhysRevLett.85.5468,PhysRevE.66.035101}.
In particular, previous studies have shown that removing high centrality nodes is an effective strategy~\citep{kitsak2010identification}.
However, deleting high centrality nodes in the network means deleting influential users on social media.
Therefore, such methods are considered to be costly~\citep{ZAREIE2021103094} and may even have a considerably negative effect because famous celebrities may be removed.
%However, deleting high-out-degree nodes in the network means deleting influential users on social media.
%Therefore, such methods are considered to be costly~\citep{kitsak2010identification} and may even have a considerably negative effect because famous celebrities may be removed.

Deleting a certain number of links in a social network is another approach to limiting the spread of information diffusion~\citep{tong2012gelling,yan2019rumor,ZAREIE2022100206,10.1007/978-3-662-47401-3_9,Kanwar2022}.
Here, deleting a link between two nodes means blocking information diffusion between them.
\cite{tong2012gelling} addressed the NetMelt problem of selecting links for deletion in order to minimize the spread of rumors.
Because the largest eigenvalue of the adjacency matrix of a social network graph is positively correlated with the size of information diffusion on the network, they proposed an algorithm to remove $ k $ links so that the largest eigenvalue of the network is minimized.
\cite{yan2019rumor} proposed a rumor spread minimization (RSM) problem and its algorithms.
RSM algorithms aim to identify a set of links so that the size of the information diffusion cascade from given source nodes is minimized.
In contrast, \cite{ZAREIE2022100206} proposed a source-ignorant method to identify a set of critical links for limiting the sizes of information diffusion cascades from arbitrary seed nodes.
These studies used synthetic information diffusion models and demonstrated the potential of link deletion methods in limiting the size of information diffusion.
Our study follows and extends these studies, evaluating the
effectiveness of link deletion methods by using actual logs of retweet
cascades on social media rather than using synthetic diffusion models.

\section{Method of Estimating Size of Diffusion Cascades after Link Deletion}
\label{sec:diffusion}
We propose a method to estimate the size of a given information diffusion cascade when deleting several links in the social network.
Notations used in this paper are summarized in Tab.~\ref{tab:notations}.
Let $T$ be a set of tweets or topics posted on social media during a certain period of time, and let $L$ be a set of links to be deleted.
Here, deleting link $(u,v) \in L$ means blocking information diffusion from user $u$ to user $v$.
We consider a method to estimate the size of each diffusion cascade regarding tweet or topic $t \in T$ after deleting links in $L$. Although each $t$ can be either a specific tweet or a specific topic, from here on, we use tweet $t$ rather than tweet or topic $t$ for simplicity.
We denote the set of users who post or repost tweet $t$ as $R_t$, and the set of users involved in tweet set $T$ as $V= \bigcup_{t \in T} R_t$.
The time at which user $u \in V$ posts $t$ or repost $t$ is denoted as $\tau(u,t)$, and the social network that expresses who-follows-whom relationships among users belonging to user set $V$ is denoted as $G=(V,E)$.
The link $(u,v) \in E$ represents user $u$ following user $v$.

\begin{table}[t]
  \begin{center}
    \caption{Notations used in this paper}
   \resizebox{.99\columnwidth}{!}{\begin{tabular}{c|c} \toprule
    Notation & Description \\ \hline
    $ G =(V,E)$ & social network with node set $V$ and link set $E$\\
    $ T $ & set of tweets or topics during a certain period\\
    $ R_t $ & set of users involved in tweet or topic $t$\\
    $ \tau (u,t)$ & time when user $ u $ posts a tweet or retweet about $t$ \\
    $ (u,v) \in E $ & link representing user $u$ following user $v$ \\
    $ H_t = (R_t,E_t)$ & diffusion graph with node set $R_t$ and link set $E_t$\\
    $ E_t $ & set of links in diffusion graph $H_t$ \\
    $ L $ & set of links to be deleted \\
    $ k $ & number of links to be deleted \\
    $ H'_t $ & diffusion graph after link deletion\\
    $ E'_t $ & set of links after link deletion in diffusion graph $H'_t$\\
    \bottomrule
  \end{tabular}}
  \label{tab:notations}
\end{center}
\end{table}

First, for each tweet $t \in T$, we construct a diffusion graph $H_t=(R_t,E_t)$ that represents the diffusion paths of tweet $t$.
The set of nodes in $H_t$ is equal to $R_t$, which is the set of users who tweeted or retweeted tweet $t$.
The link $(u,v) \in E_t$ in $H_t$ represents tweet $t$ spreading from user $u$ to user $v$.
To obtain $E_t$ is not obvious because each cascade does not contain
information about the specific diffusion paths of the tweet.  Here, we
construct three diffusion graphs, one of which is non-tree diffusion
graph (i.e., each node can have multiple parents), and the others are
trees (i.e., each node has at most a single parent).

In the non-tree diffusion graph, we consider a tweet to have spread from user $u$ to user $v$ when user $v$ is following user $u$ and the timing of user $u$'s retweeted (or tweeted) tweet $t$ is earlier than that of user $v$.
In other words, if $(v,u) \in E$ and $\tau(u,t) < \tau(v,t)$, then $(u,v) \in E_t$.
%We call the graph constructed in this way non-tree.

While a user can be affected by multiple users in the non-tree diffusion
graph, a user is affected by a only single user in the diffusion trees.  We
construct two-types of diffusion trees, which we call diffusion tree (first) and diffusion tree (last).  In the diffusion tree (first), we consider a tweet to have spread from user $u$ to user $v$ when user $v$ is following user $u$ and the timing of user $u$'s retweeted tweet $t$ is earlier than that of user $v$ and the earliest among user $v$'s followees.  In contrast, in the diffusion tree (last) we consider a tweet to have spread from user $u$ to user $v$ when user $v$ is following user $u$ and the timing of user $u$'s retweeted tweet $t$ is earlier than that of user $v$ and the latest among user $v$'s followees.
% In the non-tree diffusion graph, if a user is influenced by more than one user, it is constructed based on assumption that the information was received from all of them. This may lead to an overestimation of the number of users who received the information.
% In the diffusion graph, we can estimate those who received the
% information more positive.

We call the node that does not have any incoming links in $H_t$ the seed user of tweet $t$.
The user who posted tweet $t$ is the seed user.
User $u$ who retweeted tweet $t$ is also a seed user if none of their followees retweeted tweet $t$ before user $u$ retweeted tweet $t$.
We denote the set of seed users of tweet $t$ as $S_t$.

Next, we use the link deletion method to select a set of links $L$ ($|L|=k$) to be deleted and construct a diffusion graph after link deletion. By removing these links from diffusion graph $H_t$, we can estimate how the number of users who receive each tweet $t \in T$ changes.
We remove the selected link set $L$ from $H_t$.
Namely, we construct $H'_t=(R_t,E'_t)$, where $E'_t=E_t \setminus L$.
Deleting link $(u,v)$ corresponds to blocking information propagation from user $u$ to user $v$.
Therefore, if user $v$ receives tweet $t$ from only user $u$, that is, if node $v$ has no incoming link in $H'_t$, then deleting link $(u,v)$ will prevent user $v$ from receiving tweet $t$.

Finally, we use diffusion graph $H'_t$ to find the number of users who receive tweet $t$ after the link deletion.
In graph $H'_t$, seed users can receive tweet $t$ even after the link deletion.
Users who follow the seed users can also receive this information.
Furthermore, users who follow those users (i.e., the users following the seed users) can also receive the information.
Therefore, we assume that all users within reach of the seed users in diffusion graph $ H'_t  $ can receive tweet $t$.
Then, we estimate the number of users who receive tweet $t$ after link deletion as the number of nodes that are within the reach of the seed nodes $s \in S_t$ in diffusion graph $ H'_t $

Note that there is a only single diffusion path for a given node in the diffusion
trees whereas there are multiple diffusion paths in the non-tree diffusion
graph.  Hence, when using diffusion trees, the estimated sizes of
diffusion cascades after link deletion are smaller than those when using
non-tree diffusion graphs.  Using the non-tree diffusion graph and
diffusion tress, we examine the effectiveness of link deletion both
under the pessimistic and optimistic scenarios.

\begin{figure}[tb]
 \begin{minipage}[t]{.99\columnwidth}
\centering
\includegraphics[scale=0.1]{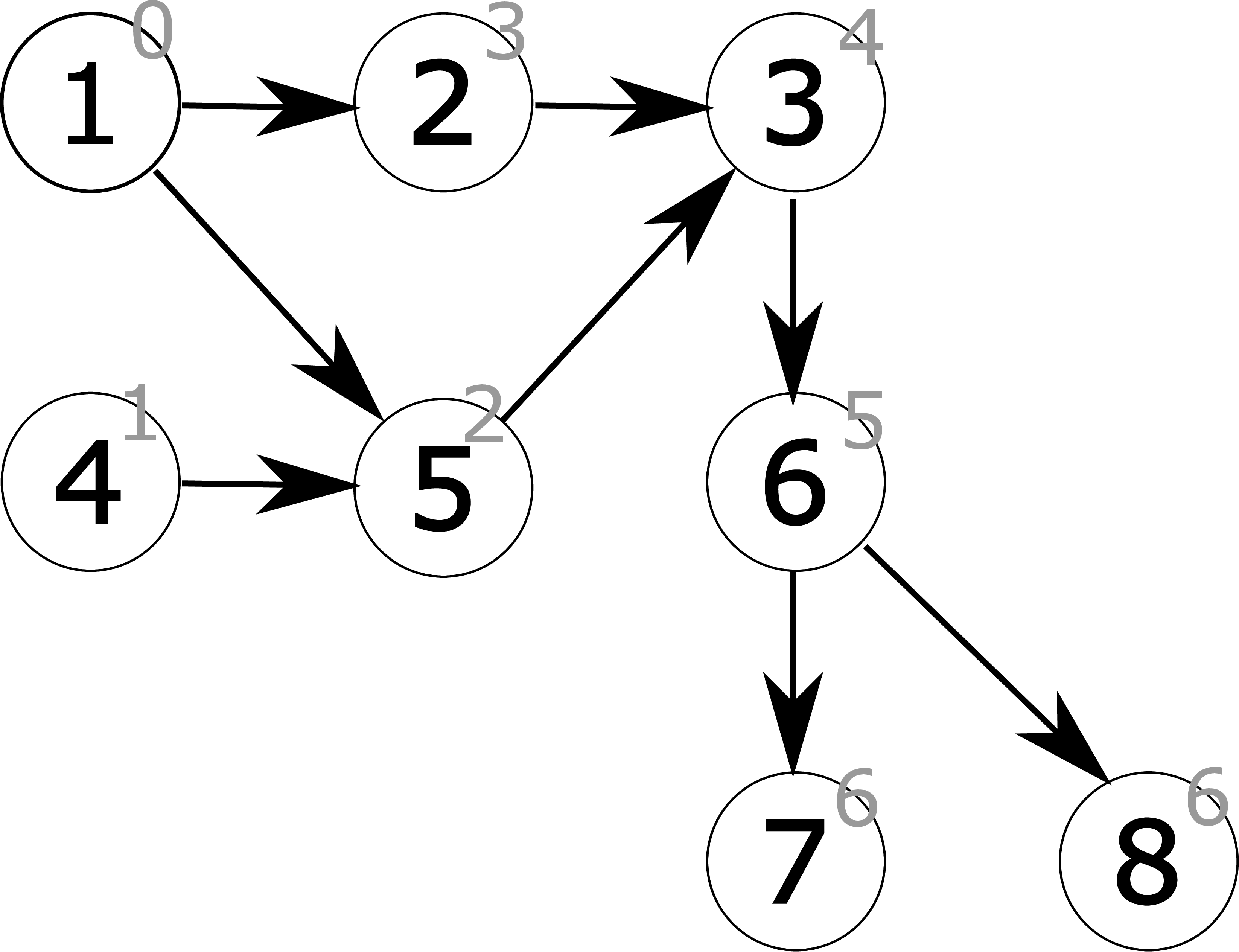}
\subcaption{Example of non-tree diffusion graph $H_t=(R_t,E_t)$ of tweet $t$. The set of users who have tweeted or retweeted tweet $t$ is $ R _t = \{ 1,2,3,4,5,6,7,8\} $, and the set of links is $ E _t = \{ (1,2),(1,5),(2,3),(4,5),(5,3),(3,6),
(6,7),(6,8) \} $. The seed node set that does not have any incoming links is $ S _t = \{ 1,4 \}$.}\label{fig:ex_graph}
\end{minipage} \\
\begin{minipage}[t]{.99\columnwidth}
\centering
\includegraphics[scale=0.1]{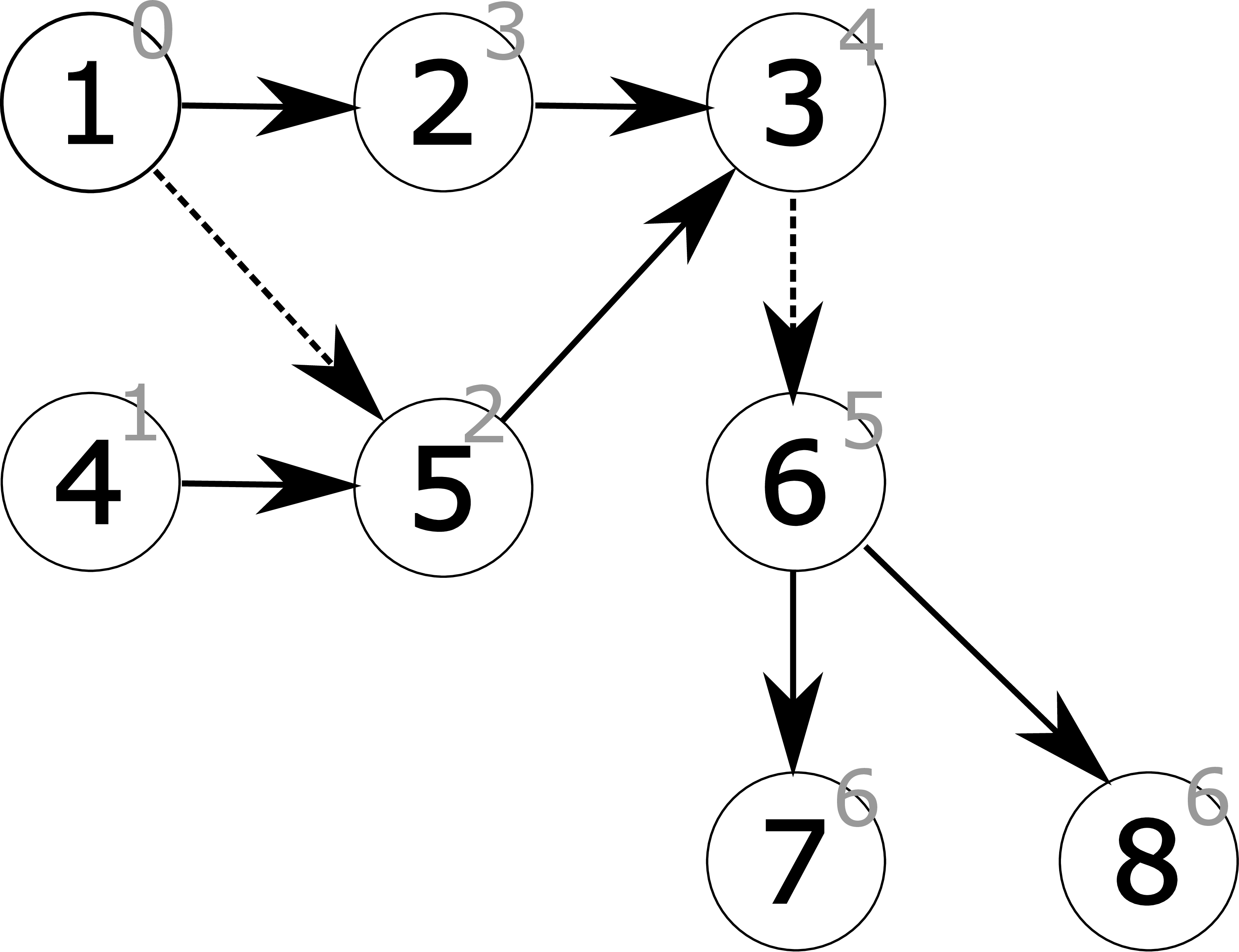}
\subcaption{Graph $ H' _t $ after deleting links $ L = \{ (1,5),(3,6) \}$ from $ H _t $. In $ H'_t $, node 6 has no incoming links, and therefore user 6 will not receive tweet $t$ that was originally tweeted by users 1 and 4. Moreover, users 7 and 8 who received tweet $t$  from user 6 also will not receive tweet $t$. The nodes within the reach of the seed node set $ S _t = \{ 1,4 \} $ are $ \{ 1,2,3,4,5 \} $, which is the set of users who will receive tweet $ t $ after the link deletion.}\label{fig:ex_delete_graph}
\end{minipage}
\caption{Examples of original diffusion graph $H_t$ and the diffusion graph after link deletion $H'_t$. Numbers on the nodes indicate the time at which the user posted a tweet or retweet regarding $t$. Link $ (1,2) $ indicates that user 2 is following user 1, and the time at which user 1 posted a tweet or retweet regarding $t$ was earlier than the time at which user 2 did.}
\end{figure}

An example of non-tree diffusion graph $H_t=(R_t,E_t)$ of tweet $t$ is shown in Fig.~\ref{fig:ex_graph}.
We consider the case where the set of links to be deleted is $ L = \{ (1,5),(3,6) \}$.
Graph $ H' _t $ after deleting these links from $ H _t $ is shown in Fig.~\ref{fig:ex_delete_graph}.

\section{Dataset and Methodology}
\label{sec:method}

\subsection{Datasets}
To evaluate the effectiveness of link deletion in limiting the size of cascades, we use four datasets containing both cascades and social networks representing who-follows-whom relationships among users involved in the cascades.
A statistical summary of the datasets is shown in Tab.~\ref{tab:dataset}. The datasets are explained in detail as follows.
\begin{enumerate}
\item Ordinary dataset~\citep{tsugawa2019empirical}

This dataset contains 10,000 tweets and their retweets posted during November 19--25, 2018.
The 10,000 tweets were randomly selected from a collection of English tweets with one or more retweets posted during the above period.
Note that there were no specific events during the period.
In total, 114,249 users tweeted or retweeted these tweets, and these users are target user set $V$.
This dataset also contains the social network of who-follows-whom relationships among the target users.
The social network is used to construct the diffusion graph and to determine links to be deleted.
To examine the effectiveness of link deletion in limiting the size of retweet cascades, tweets that have a large cascade are preferable for analysis.
Therefore, among the 10,000 tweets, we extracted those that were retweeted 100 times or more, which resulted in 90 original tweets. These 90 tweets that were retweeted 100 times or more are the set of tweets $T$ that we analyzed.

\item Higgs dataset~\citep{de2013anatomy}

This dataset contains tweets and retweets about the discovery of the Higgs boson in 2012.
In this dataset, the discovery of the Higgs boson is regarded as topic $t$.
That is, all sets of tweets and retweets in the dataset are regarded as a single cascade of information diffusion.
Therefore, the number of topics to be analyzed $T=1$ in this dataset.
The social network representing who-follows-whom relationships among users who tweet and retweet is available in this dataset and is used to construct the diffusion graph and to determine links to be deleted.

\item URL dataset~\citep{hodas2014simple}

This dataset contains tweets containing URLs and retweets during October 2010.
Each URL is considered as a topic propagated among users.
As with the ordinary dataset, we extract tweets that were retweeted 100 times or more.

\item Douban dataset~\citep{10.1145/2339530.2339641}

This dataset contains data from a Chinese social website.
This website allows users to post their book reading status and also allows them to follow the status of other users.
Each book is considered as a topic propagated among users.
As with the ordinary dataset, we extract cascades whose size were 100 or more.

\begin{table}[t]
  \begin{center}

    \caption {Statistics of the datasets used in the analysis}
    \resizebox{.95\linewidth}{!}{
  \begin{tabular}{c|c|c|c|c} \toprule
    & Ordinary & Higgs & Douban & URL\\ \hline
    Number of target users $ V $ &118,162&456,626&25,306&12,627\\
    Number of links $ E $  &3,130,963&14,855,842&758,310&619,262\\
    Number of target tweets or topics $ T $ &90&1&439&253\\
    Average cascade size &830.4&228,556&272.6&246.5\\
\bottomrule
      \end{tabular}
      }
      \label{tab:dataset}
\end{center}
\end{table}

\end{enumerate}

\subsection{Link Deletion Methods}

In this study, we focus on the effectiveness of three popular link
deletion methods, NetMelt~\citep{tong2012gelling}, Betweenness~\citep{KASHYAP2019631}, and Edge-Degree~\citep{KASHYAP2019631} in limiting the
size of actual diffusion cascades. We also use a random baseline. The
details of the link deletion methods are explained as follows.

\begin{enumerate}
  \item NetMelt~\citep{tong2012gelling}

  We apply the NetMelt to the social network $G$ representing
	who-follows-whom relationships among the target users and obtain
	$|L|=k$ links to be deleted.

 \item Betweenness~\citep{KASHYAP2019631}

 Betweenness selects $|L|=k$ links to be deleted in descending order of
       link betweenness of the links in social network $G$.

 \item Edge Degree~\citep{KASHYAP2019631}

 The edge degree of link $(u,v)$ is defined as the product of in-degree
       of node $u$ and out-degree of node $v$~\citep{KASHYAP2019631}.
This method selects $|L|=k$ links to be deleted in descending order of
       edge degree of the links in social network $G$.

  \item Random

  Random selects $|L|=k$ links to be deleted randomly from all links in social network $G$.

\end{enumerate}

Using these methods, we determine $|L|=k$ links to be deleted and estimate the size of cascade $t \in T$ after link deletion.

\section{Results and Discussion}
\label{sec:result}

\begin{figure}[tbp]
\begin{tabular}{cc}
\begin{minipage}[t]{0.49\columnwidth}
  \begin{center}
    \includegraphics[keepaspectratio,scale=0.6]{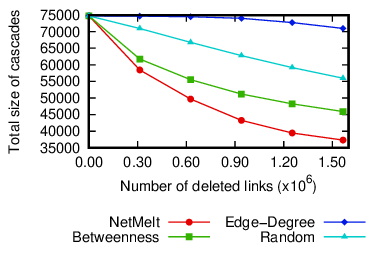}
    \subcaption{Ordinary}
    \label{fig:non_tree_orig}
  \end{center}
\end{minipage} &
  \begin{minipage}[t]{0.49\columnwidth}
    \begin{center}
      \includegraphics[keepaspectratio,scale=0.6]{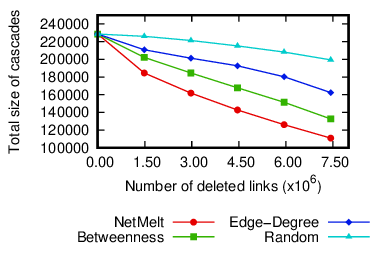}
      \subcaption{Higgs}
      \label{fig:non_tree_higgs}
    \end{center}
  \end{minipage}  \\
  \begin{minipage}[t]{0.49\columnwidth}
    \begin{center}
      \includegraphics[keepaspectratio,scale=0.6]{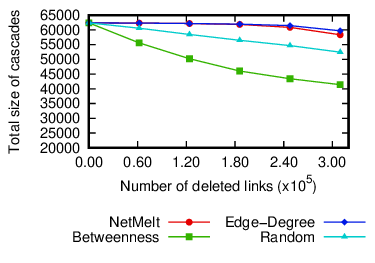}
      \subcaption{URL}
      \label{fig:non_tree_twitter}
    \end{center}
    \end{minipage} &
    \begin{minipage}[t]{0.49\columnwidth}
      \begin{center}
        \includegraphics[keepaspectratio,scale=0.6]{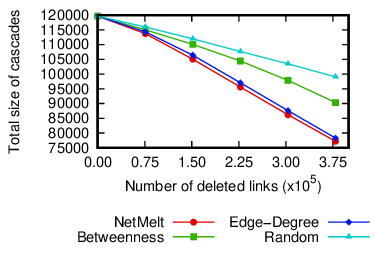}
        \subcaption{Douban}
        \label{fig:non_tree_douban}
      \end{center}
  \end{minipage}
  \end{tabular}
\caption{Total size of cascades after link deletion vs. the number of deleted links (non-tree diffusion graph)}
\label{fig:non_tree}
\end{figure}

First, we analyze how the total size of the cascades in the four datasets changes by link deletion.
Figure~\ref{fig:non_tree} shows the estimated total size of retweet
cascades against the number of deleted links in each dataset when using
non-tree diffusion graph. From Fig.~\ref{fig:non_tree}, we can see that
although the total cascade size decreases by link deletion, the decrease
is not large relative to the number of deleted links. For example, in
the case of ordinary data, when 1.5 million links are deleted from the
social network using NetMelt, the total cascade size is about 50\% that
of the original cascade. 1.5 million links are equivalent to
approximately 50\% of the links in the original social network. It is
practically difficult to block information diffusion between such a
large number of user pairs. Therefore, this result suggests that the
effects of link deletion methods to limit the size of tweet diffusion cascades are limited.
Looking at the difference among link deletion methods, the NetMelt and
the Betweenness are generally effective.  In particular, except for the
URL dataset, the NetMelt
achieves the best results in the three of four datasets.

\begin{figure}[tbp]
\begin{tabular}{cc}
\begin{minipage}[t]{0.49\columnwidth}
  \begin{center}
    \includegraphics[keepaspectratio,scale=0.6]{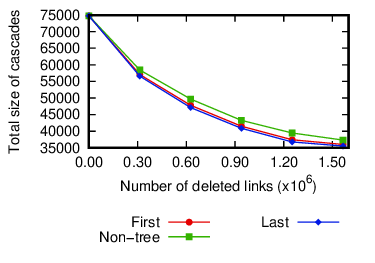}
    \subcaption{Ordinary}
    \label{fig:netmelt_orig}
  \end{center}
\end{minipage} &
  \begin{minipage}[t]{0.49\columnwidth}
    \begin{center}
      \includegraphics[keepaspectratio,scale=0.6]{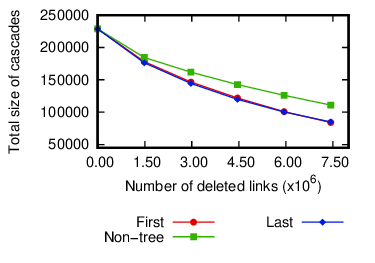}
      \subcaption{Higgs}
      \label{fig:netmelt_higgs}
    \end{center}
  \end{minipage}  \\
  \begin{minipage}[t]{0.49\columnwidth}
    \begin{center}
      \includegraphics[keepaspectratio,scale=0.6]{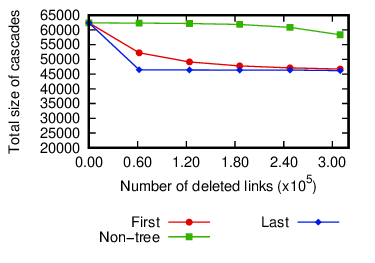}
      \subcaption{URL}
      \label{fig:netmelt_twitter}
    \end{center}
    \end{minipage} &
    \begin{minipage}[t]{0.49\columnwidth}
      \begin{center}
        \includegraphics[keepaspectratio,scale=0.6]{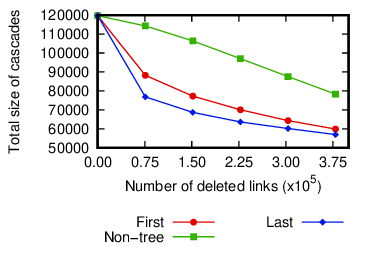}
        \subcaption{Douban}
        \label{fig:netmelt_douban}
      \end{center}
  \end{minipage}
  \end{tabular}
\caption{Comparison among different diffusion graphs (NetMelt)}
\label{fig:netmelt}
\end{figure}

\begin{figure}[tbp]
\begin{tabular}{cc}
\begin{minipage}[t]{0.49\columnwidth}
  \begin{center}
    \includegraphics[keepaspectratio,scale=0.6]{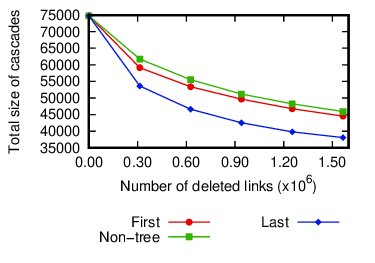}
    \subcaption{Ordinary}
    \label{fig:betweenness_orig}
  \end{center}
\end{minipage} &
  \begin{minipage}[t]{0.49\columnwidth}
    \begin{center}
      \includegraphics[keepaspectratio,scale=0.6]{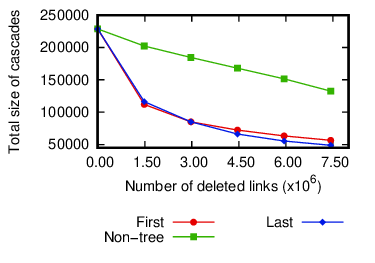}
      \subcaption{Higgs}
      \label{fig:betweenness_higgs}
    \end{center}
  \end{minipage}  \\
  \begin{minipage}[t]{0.49\columnwidth}
    \begin{center}
      \includegraphics[keepaspectratio,scale=0.6]{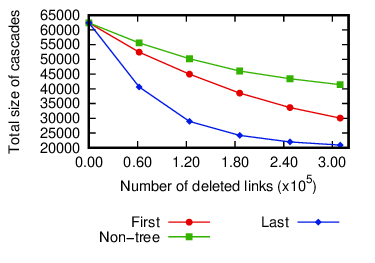}
      \subcaption{URL}
      \label{fig:betweenness_twitter}
    \end{center}
    \end{minipage} &
    \begin{minipage}[t]{0.49\columnwidth}
      \begin{center}
        \includegraphics[keepaspectratio,scale=0.6]{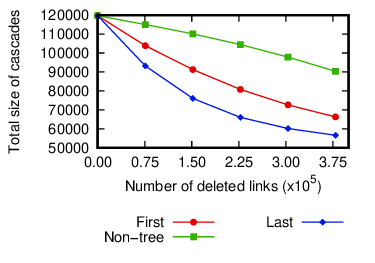}
        \subcaption{Douban}
        \label{fig:betweenness_douban}
      \end{center}

  \end{minipage}
  \end{tabular}
\caption{Comparison among different diffusion graphs (Betweenness)}
\label{fig:betweenness}
\end{figure}

We next examine the effectiveness of link deletion under more optimistic
scenarios by using diffusion trees.
Figures~\ref{fig:netmelt} and~\ref{fig:betweenness} compare the total
size of cascades when using non-tree diffusion graph, diffusion tree
(first), and diffusion tree (last).  Figure~\ref{fig:netmelt} shows the
results for the NetMelt, and Fig.~\ref{fig:betweenness} shows the results
for the Betweenness.  These results show that although the estimated
sizes of cascades when using the diffusion trees are smaller than those when
using the non-tree diffusion graph, the estimated sizes of cascades are
still large.  Even under the best
case for each dataset, 10\%--50\% links are necessary to be deleted for
limiting the estimated diffusion sizes to be under 50\% of the original
sizes.
These results again suggest the limitations of link deletion methods for
limiting the sizes of real diffusion cascades.  It is suggested that even under the
optimistic estimation, many links should be deleted to reduce the
cascade sizes.

% Next, we examine differences among link deletion methods. When considering the difference between NetMelt and RTNetMelt (Fig. 3a), we find that NetMelt is considerably more effective compared with RTNetMelt. Moreover, we also find that RTNetMelt is less effective than the Random baseline. This suggests that using a retweet network is not a good option, and the social network representing who-follows-whom relationships should be used when applying the NetMelt algorithm. The reason for the poor effectiveness of RTNetMelt is unclear from the results, and further experiments will be necessary to clarify the reason.

\begin{figure}[t]
\begin{tabular}{cc}
 \begin{minipage}[t]{0.45\linewidth}
  \begin{center}
  \includegraphics[keepaspectratio,scale=0.6]{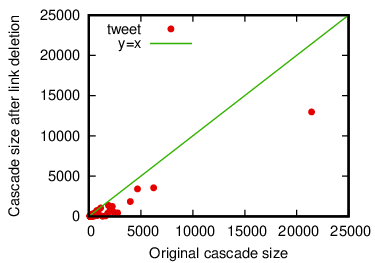}
  \subcaption{Ordinary}\label{fig:scat-orig-netmelt}
  \end{center}
  \end{minipage} &
   \begin{minipage}[t]{0.45\linewidth}
  \begin{center}
  \includegraphics[keepaspectratio,scale=0.6]{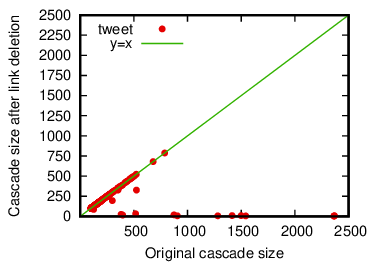}
  \subcaption{URL}\label{fig:scat-twitter-netmelt}
  % \includegraphics[keepaspectratio,scale=0.6]{figure/result_figure/orig/scat_rand_orig_50_last.eps}
  % \subcaption{Random}\label{fig:scat-orig-rand}
  \end{center}
  \end{minipage} \\
 \begin{minipage}[t]{0.45\linewidth}
  \begin{center}
  \includegraphics[keepaspectratio,scale=0.6]{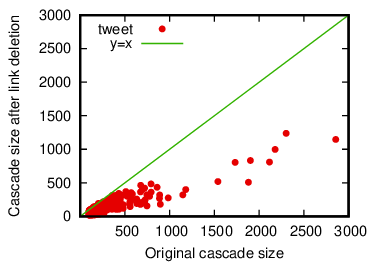}
  \subcaption{Douban}\label{fig:scat-douban-netmelt}
  % \includegraphics[keepaspectratio,scale=0.6]{figure/result_figure/orig/scat_betweenness_orig_50_last.eps}
  % \subcaption{Betweenness}
  % \label{fig:scat-orig-bet}
  \end{center}
  \end{minipage} &
  %  \begin{minipage}[t]{0.45\linewidth}
  % \begin{center}
  % \includegraphics[keepaspectratio,scale=0.6]{figure/result_figure/orig/scat_degree_orig_50_last.eps}
  % \subcaption{Degree}
  % \label{fig:scat-orig-degree}
  % \end{center}
  % \end{minipage}
  \end{tabular}
  \caption{Relation between the original cascade size and the cascade
 size after link deletion for each tweet (link deletion method: NetMelt,
 diffusion graph: diffusion tree (last), number of deleted links $|L|=0.5|E|$)}
\label{fig:scatter-netmelt}
\end{figure}

\begin{figure}[t]
\begin{tabular}{cc}
 \begin{minipage}[t]{0.45\linewidth}
  \begin{center}
  \includegraphics[keepaspectratio,scale=0.6]{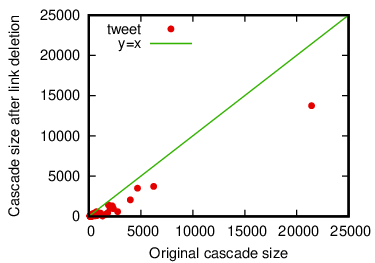}
  \subcaption{Ordinary}\label{fig:scat-orig-betweenness}
  \end{center}
  \end{minipage} &
   \begin{minipage}[t]{0.45\linewidth}
  \begin{center}
  \includegraphics[keepaspectratio,scale=0.6]{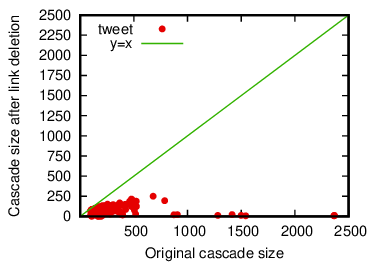}
  \subcaption{URL}\label{fig:scat-twitter-betweenness}
  \end{center}
  \end{minipage} \\
 \begin{minipage}[t]{0.45\linewidth}
  \begin{center}
  \includegraphics[keepaspectratio,scale=0.6]{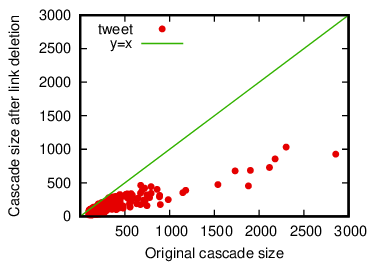}
  \subcaption{Douban}\label{fig:scat-douban-betweenness}
  \end{center}
  \end{minipage} &

  \end{tabular}
  \caption{Relation between the original cascade size and the cascade
 size after link deletion for each tweet (link deletion method: Betweenness,
 diffusion graph: diffusion tree (last), number of deleted links $|L|=0.5|E|$)}
\label{fig:scatter-bet}
\end{figure}

We next analyze how the link deletion affects the size of each
cascade. Because the Higgs dataset contains only a single cascade, the
other three datasets are used in the following
analyses. Figures~\ref{fig:scatter-netmelt} and~\ref{fig:scatter-bet}
show the relation between the original cascade size and the estimated
cascade size after link deletion for each tweet when using the NetMelt
and the Betweenness, respectively. The number of deleted
links is 50\% of the original links, and the diffusion tree (last) is
used as the diffusion graph.
These results show that the cascade size of some tweets is substantially decreased by link deletion.
However, we should note that this result is obtained when 50\% of links are deleted.
Considering the fact that a large number of links are deleted, this result also suggests that the effect of link deletion on information diffusion is limited.

% \begin{figure}[t]
%  \begin{minipage}[t]{0.45\linewidth}
%   \begin{center}
%   \includegraphics[keepaspectratio,scale=0.6]{figure/result_figure/orig/scat_NetMelt_30_all.eps}
%   \subcaption{All tweets}\label{fig:scat-all}
%   \end{center}
%   \end{minipage}
%  \begin{minipage}[t]{0.45\linewidth}
%   \begin{center}
%   \includegraphics[keepaspectratio,scale=0.6]{figure/result_figure/orig/scat_NetMelt_30_1000.eps}
%   \subcaption{Tweets with less than 1000 retweets}
%   \label{fig:scat-1000}
%   \end{center}
%   \end{minipage}
%   \caption{Relation between the original cascade size and the cascade size after link deletion for each tweet (dataset: ordinary dataset, link deletion method: NetMelt, number of deleted links $|L|=300,000$)}
% \label{fig:scatter-RT}
% \end{figure}

\begin{figure}[t]
\begin{tabular}{cc}
 \begin{minipage}[b]{0.45\linewidth}
  \begin{center}
  \includegraphics[keepaspectratio,scale=0.6]{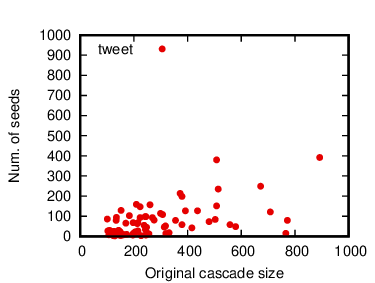}
  \subcaption{Ordinary}
  \label{fig:scat-seed-1000}
  \end{center}
  \end{minipage} &
   \begin{minipage}[b]{0.45\linewidth}
  \begin{center}
  \includegraphics[keepaspectratio,scale=0.6]{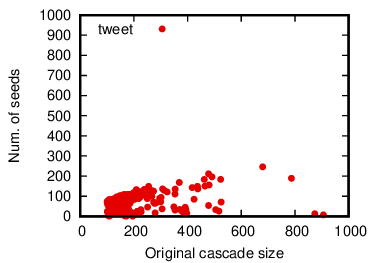}
  \subcaption{URL}
  \end{center}
  \end{minipage} \\
 \begin{minipage}[b]{0.45\linewidth}
  \begin{center}
  \includegraphics[keepaspectratio,scale=0.6]{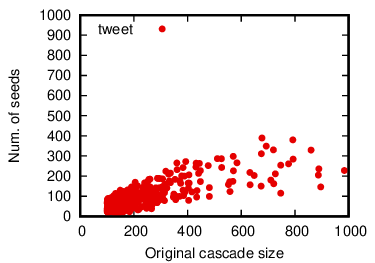}
  \subcaption{Douban}
  \end{center}
  \end{minipage} &
  \end{tabular}
  \caption{Relation between the original cascade size and the number of
 seed nodes for each tweet (tweets with less than 1000 retweets)}
\label{fig:scat-seed}
\end{figure}

Finally, we examine the reason why the link deletion does not substantially reduce the size of diffusion cascades.
Figure~\ref{fig:scat-seed} shows the relationship between the original
cascade size of each tweet and the number of seed users of each tweet.
For the sake of visibility, we only show the tweets with less than 1000 retweets.
These figures show that many tweets have a large ratio of seed users relative to the total cascade size.
A seed user is a user who retweets a tweet without being influenced by their followers.
Twitter users may obtain information about tweets from sources other than their followers.
For example, if a tweet is {\em trending} on Twitter, or if it is introduced on another web page, Twitter users may obtain information about the tweet not from their social relationships on Twitter.
Link deletion only limits information diffusion through social relationships and cannot prevent seed users from obtaining information from other sources.
If a tweet spreads through only social relationships among users, link deletion should be effective in limiting its cascade size. However, if many users obtain information about a tweet from sources other than their social relationships, link deletion may not be effective in limiting its cascade size.

\begin{figure*}[t]
%1
\begin{minipage}[t]{.99\columnwidth}
 \begin{center}
 \includegraphics[width=.99\columnwidth]{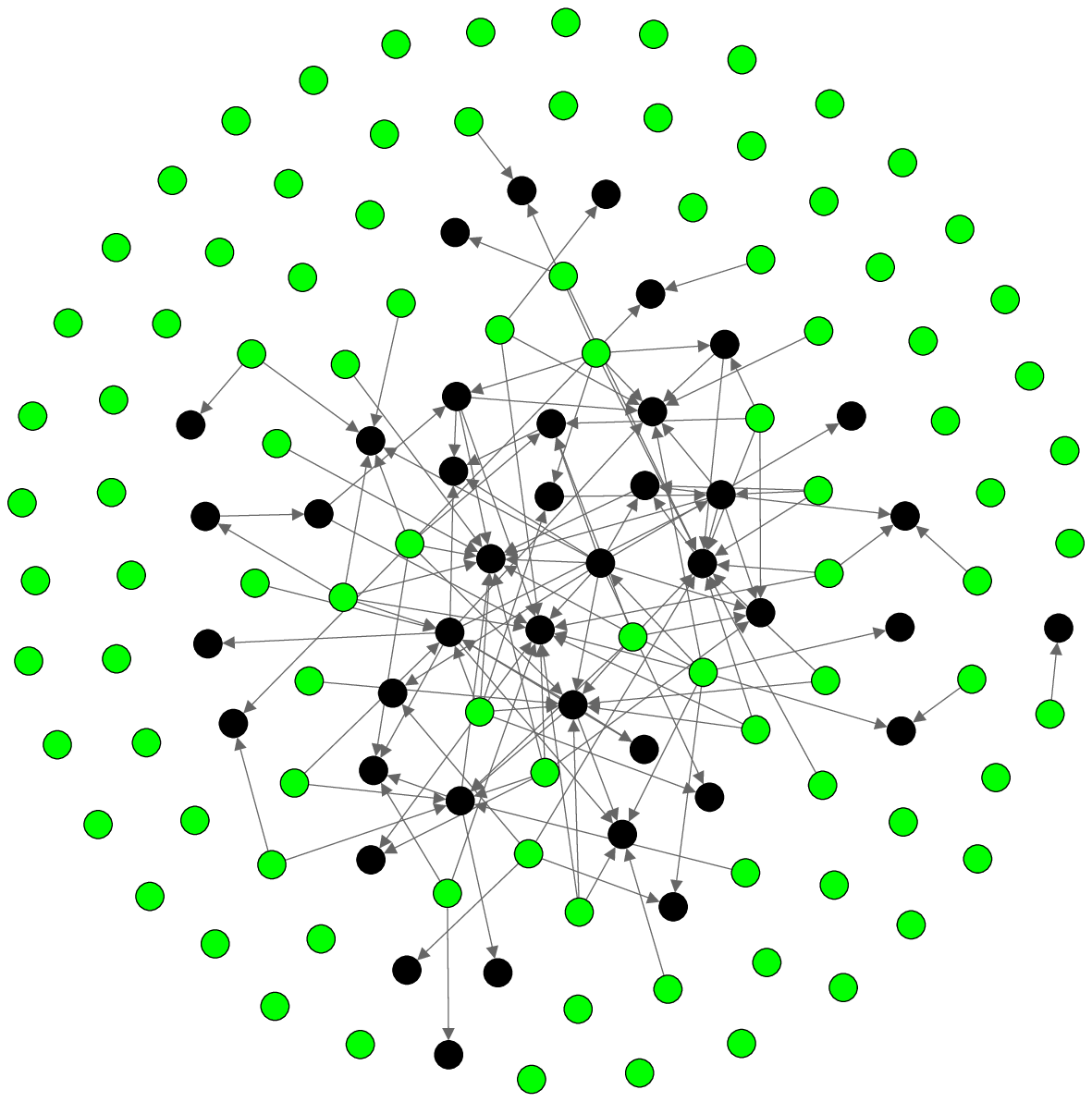}
 \subcaption{Example with a relatively large number of seed users}\label{fig:many}
 \end{center}
 \end{minipage}
 %2
 \begin{minipage}[t]{.99\columnwidth}
  \begin{center}
  \includegraphics[width=.99\columnwidth]{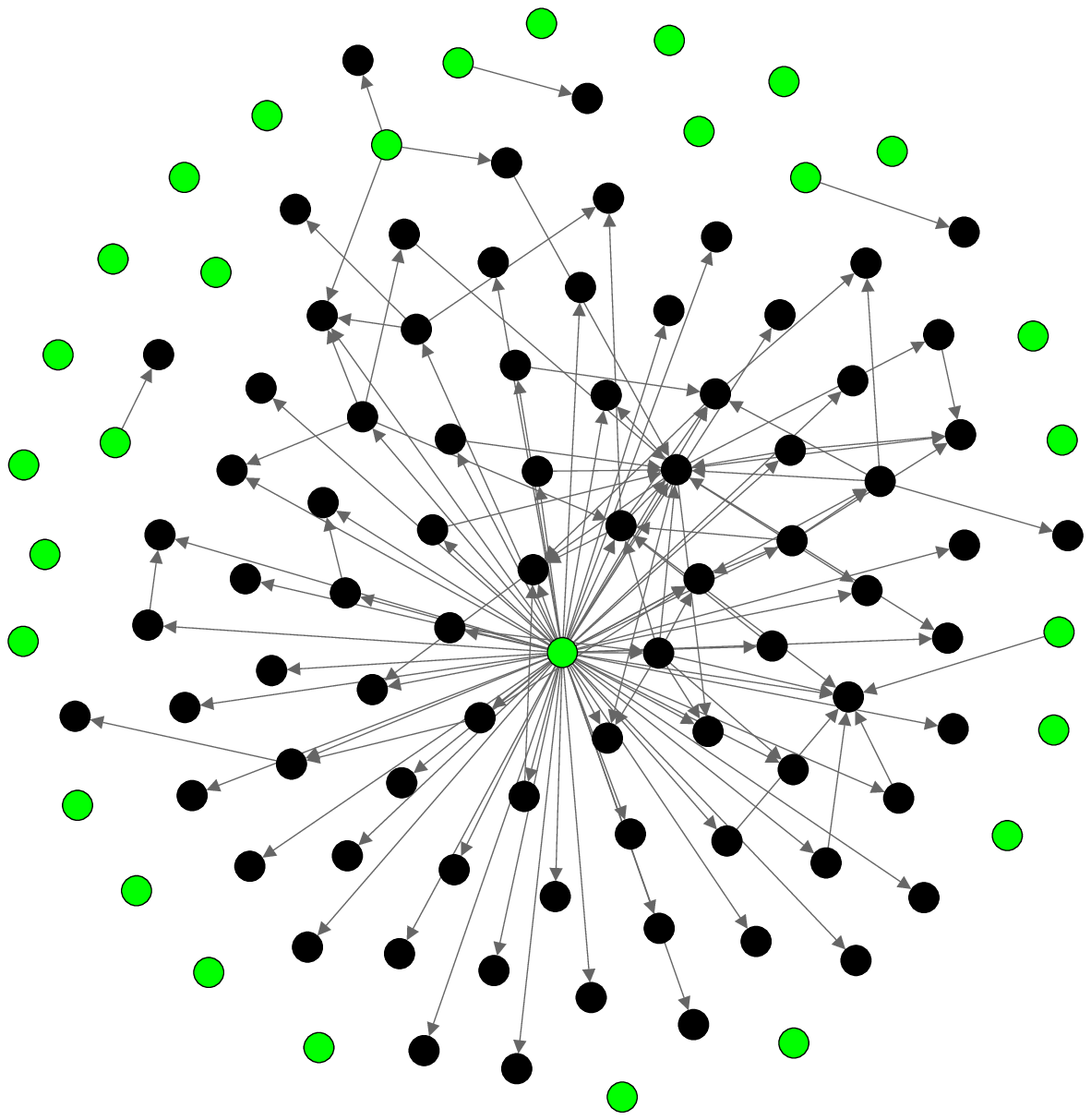}
  \subcaption{Example with a relatively small number of seed users}\label{fig:few}
  \end{center}
  \end{minipage}
  \caption{Examples of non-tree diffusion graph (ordinary dataset): Light green nodes represent seed users.}
\label{fig:ex_diffusion_graph}
\end{figure*}

From the examples of diffusion graphs shown in Fig.~\ref{fig:ex_diffusion_graph},  we can also confirm that limiting the cascade sizes by removing a small number of links is difficult.
Seed users are represented as light green nodes in the figure.
If seed users will be isolated from other users in the diffusion graph, the cascade size can be reduced.
However, in the example of Fig.~\ref{fig:many}, many seed users involve in the tweet diffusion, and isolating these seed users requires many links to be deleted.
Also for the example of Fig.~\ref{fig:few}, a seed user shown in the center of the figure is connected to many other users, and isolating the seed user requires many links to be deleted.

In summary, our results suggest that link deletion methods are limited
in efficiently limiting the size of diffusion cascades. NetMelt has been
shown to be effective in limiting the size of synthetic cascades
in~\citep{tong2012gelling} and requires many links (e.g., 50\% of links)
to be deleted to limit the size of real tweet diffusion cascades. In
practice, it is difficult to block information diffusion between such a
large number of user pairs on social media, which suggests that link
deletion on social media is limited in practice.
Our results also suggest that the existence of many seed users is one of the
reasons for the inefficiency of link deletion methods in actual social
media. We show that the information cascades analyzed in this paper
have many seed users who may receive the information, not from their
social relationships, but from other sources. Link deletion methods
cannot prevent seed users from receiving information, so
other strategies (e.g., advertising counter campaigns~\citep{Budak11:limiting,Erd20:Blocking}) are suggested as necessary methods for limiting the size of
diffusion cascades with many seed users.
Moreover, we recognize
that other link deletion methods that delete links by using the
information about the seed nodes
are available in the literature~\citep{yan2019rumor}, and their effectiveness should be evaluated using actual diffusion cascade logs in future research.

\section{Conclusion and Future Work}
\label{sec:conclusion}
In this study, we used actual logs of retweet cascades and evaluated the effectiveness of link deletion methods to suppress the spread of information.
We proposed a method to estimate the size of a given retweet cascade after link deletion and examined the potential and limitations of link deletion methods when applying them to actual social media.
Our results suggest that link deletion methods are limited in limiting
the size of retweet cascades on social media.
We showed that most of the retweet cascades have multiple sources, which makes limiting their size by link deletion difficult.

In future work, we plan to evaluate the effectiveness of link deletion
methods that incorporate the information of seed nodes~\citep{yan2019rumor}. In addition, validating the proposed estimation method of cascade size is also an important future work. % In our proposed method, we assume that seed users participate in the cascade even after all the links have been deleted. This assumption might result in overestimation of the cascade size. In actual social media, some seed users may participate in a cascade because it is large. In such a case, if many links are deleted and the cascade size is reduced, the seed users may not participate in the cascade. Considering such effects, the actual effects of link deletion should be validated in future research.

%謝辞
\section*{ACKNOWLEDGMENT}
This work was partly supported by JSPS KAKENHI Grant Number 19K11917.
\bibliographystyle{spbasic}
\bibliography{myrefs}

\end{document}